\documentclass[preprint]{article}
\newcommand{\be}{\begin{equation}} \newcommand{\ee}{\end{equation}} 
\newcommand{\bea}{\begin{eqnarray}}\newcommand{\eea}{\end{eqnarray}}

\begin{document}
\title{ Constructing Exactly Solvable Pseudo-hermitian Many-particle Quantum
Systems by Isospectral Deformation}
\author{ Pijush K. Ghosh}
\date{Department of Physics, Siksha-Bhavana,\\ 
Visva-Bharati University,\\
Santiniketan, PIN 731 235, India.}
\maketitle
\begin{abstract} 
A class of non-Dirac-hermitian many-particle quantum systems admitting
entirely real spectra and unitary time-evolution is presented. These
quantum models are isospectral with Dirac-hermitian systems and are exactly
solvable. The general method involves a realization of the basic canonical
commutation relations defining the quantum system in terms of operators
those are hermitian with respect to a pre-determined positive definite
metric in the Hilbert space. Appropriate combinations of these operators result
in a large number of pseudo-hermitian quantum systems admitting entirely
real spectra and unitary time evolution. Examples of a pseudo-hermitian
rational Calogero model and XXZ spin-chain are considered.
\end{abstract}
\section{Introduction}
The (non-)hermiticity of an operator crucially depends on the choice of the
metric( or inner-product) in the Hilbert space, which has been taken as
an identity operator in the standard treatment of quantum mechanics. The
question of necessity of such a choice in formulating quantum physics is
as old as the subject itself. A renewed interest\cite{bend,ali,quasi}
has been generated over the last decade in addressing the same question in
a systematic manner. The current understanding is that a quantum system with
unbroken combined Parity(P) and Time-reversal(T) symmetry admits entirely
real spectra even though the system may be non-Dirac-hermitian.
A consistent quantum description including reality of the entire spectra and
unitary time-evolution of the non-Dirac-hermitian system is possible with
the choice of a new inner-product\cite{bend}. An alternative  description of
non-Dirac-hermitian quantum systems admitting entirely real spectra is in
terms of pseudo-hermitian operator\cite{ali,quasi}. The existence
of a positive-definite metric in the Hilbert space is again crucial in
this formalism for showing reality of the entire spectra as well as unitary
time evolution.

A large number of non-Dirac-hermitian quantum systems admitting entirely
real spectra have been found. A few prototype examples from a vast list of
such systems are contained in Refs.
\cite{bend,ali,quasi,piju1,swan,me,piju,spin,ptcsm,pkumar,europe}.
The real difficulty in giving a complete description of these systems lies
in finding the exact positive-definite metric in the Hilbert space.
It is worth emphasizing here that no expectation values of physical
observables or correlation functions can be calculated without the knowledge
of the metric in the Hilbert space. This makes a non-Dirac-hermitian quantum
system incomplete, even though the complete energy spectrum and the associated
eigenfunctions may be known explicitly. There are not many non-Dirac-hermitian
quantum systems for which the metric is known exactly and explicitly, the
scenario being even worse for such system with several degrees of freedom.

The purpose of this contribution is to present a class of exactly solvable
pseudo-hermitian many-particle quantum systems with a pre-determined metric
in the Hilbert space. The motivation behind considering a pre-determined metric
lies in the fact that it can be used to construct non-Dirac-hermitian quantum
systems with a complete description, simply by deforming known Dirac-hermitian
quantum systems. Although the non-Dirac-hermitian Hamiltonian constructed
in this way is isospectral with the corresponding Dirac-hermitian Hamiltonian,
the difference may appear in the description of different correlation
functions of these two quantum systems\cite{piju}.

The motivation behind constructing such non-Dirac-hermitian Hamiltonian
by isospectral deformation is the following.
First of all, quantum systems constructed in this way may serve as prototype
examples for testing different ideas and methods related to the subject. This
is important particularly in the context of many-particle systems, where the
number of exactly solvable models with an explicit knowledge of the metric
in the Hilbert space are very few and validity of approximate and/or
numerical methods are required to be checked before applications.
Secondly, a significant number of non-Dirac-hermitian quantum systems are known
to admit entirely real spectra for which the origin of the reality of the
spectra is not obvious. It is to be seen whether the reality of the spectra
of some of these models could be related to certain pseudo-hermitian quantum
system or not. In fact, the asymmetric $XXZ$ spin-chain\cite{xxz}, which is
relevant in the context of two species
reaction-diffusion processes and Kardar-Parisi-Zhang-type growth phenomenon,
is shown to be pseudo-hermitian following the general approach\cite{piju1}
prescribed in this article. Finally, construction of physically realizable
quantum systems is always desirable.

The general method involves a realization of the basic canonical commutation
relations defining the quantum system in terms of operators those are
hermitian with respect to a pre-determined positive definite metric
$\eta_+$ in the Hilbert space. Consequently, any Hamiltonian
that is constructed using appropriate combination of these operators is
hermitian with respect to $\eta_+$. However, in general, the same Hamiltonian
may not be Dirac-hermitian, thereby giving rise to a pseudo-hermitian
Hamiltonian. A pseudo-hermitian quantum system constructed this way
may or may not be exactly solvable. The examples considered in this article
include an exactly solvable non-Dirac-hermitian Calogero Model and
an $XXZ$ spin-chain. Both of these models are pseudo-hermitian and isospectral
with known Dirac-hermitian model. Many other interesting quantum systems
following from this construction are described in Ref. \cite{piju1}.

\section{Preliminaries and Examples }

The Hilbert space that is endowed with the standard inner product
$\langle ., . \rangle$ is denoted as ${\cal{H}}_D$. The subscript $D$
indicates that the Dirac-hermiticity condition is used in this Hilbert space.
On the other hand, the Hilbert space that is endowed with the
positive-definite metric $\eta_+$ and the modified inner product,
\be
\langle \langle . , . \rangle \rangle_{\eta_+} := \langle ., \eta_+ . \rangle.
\label{ip}
\ee
\noindent is denoted as ${\cal{H}}_{\eta_+}$. Corresponding to a hermitian
operator $\hat{\cal{B}}$ in the Hilbert space ${\cal{H}}_D$, a hermitian
operator $\hat{B}$ in the Hilbert space ${\cal{H}}_{\eta_+}$ can be
defined as\cite{ali},
\be
\hat{B} = \rho^{-1} {\hat{\cal{B}}} \rho, \ \ \ \rho:=\sqrt{\eta_+}.
\label{obs}
\ee
\noindent An interesting consequence of Eq. (\ref{obs}) is that a set of
operators $\hat{\cal{B}}_i$ obey the same canonical commutation relations
as those satisfied by the corresponding set of operators $\hat{B}_i$ and the
vice verse. The relation (\ref{obs}) is important for identifying observables
in ${\cal{H}}_{\eta_+}$ and also crucial for the discussion that follows.

\subsection{Pseudo-hermitian Rational Calogero Model}

The rational Calogero model\cite{csm} is one of the most
celebrated examples of exactly solvable many-particle quantum systems.
This model\cite{csm} and its variants\cite{poly} are relevant to the study of
a diverse branches of contemporary physics. Calogero-Sutherland-type models
have been constructed previously\cite{ptcsm,pkumar,europe} within the context
of PT-symmetric quantum systems. A new class of pseudo-hermitian quantum
system involving rational Calogero model is presented below.

A Dirac-non-hermitian rational $A_{N+1}$ Calogero model may be introduced as
follows:
\bea
H & = & - \frac{1}{2} \sum_{i=1}^N \frac{\partial^2}{\partial x_i^2}
+ \frac{1}{2} \lambda (\lambda-1) \sum_{i \neq j} X_{ij}^{-2} +
\frac{1}{2} \sum_{i=1}^N  x_i^2,\nonumber \\
X_{12} & = & \left ( x_1 - x_2 \right ) cosh \phi + i \left ( x_1 +
x_2 \right ) sinh \phi,\nonumber \\
X_{1j} & = &  x_1 cosh \phi + i x_2 sinh \phi - x_j, \ \ j > 2,\nonumber \\
X_{2 j} & = &  -i x_1 sinh \phi + x_2 cosh \phi - x_j \ \ j > 2,\nonumber \\
X_{ij} & = & x_i - x_j, \ \  (i, j) > 2.
\label{calo}
\eea
\noindent The parameters $\lambda, \phi$ appearing in $H$ are real. The
coordinates $x_i$ and their conjugate momenta $p_i$ are hermitian in 
${\cal{H}}_D$. Unlike the standard rational Calogero model, the
two-body inverse-square interaction term is neither invariant under
translation nor singular for $x_{1} = x_i, i > 1$ and  $x_2 = x_i, i > 2$.
However, the Hamiltonian is invariant under a combined $PT$ operation
with the $P$ and $T$ transformations defined as,
\be
P: x_1 \rightarrow x_2, \ \ x_2 \rightarrow x_1, \ \ x_i \rightarrow x_i \ \
\forall \ i > 2; \ \ T: i \rightarrow - i.
\label{2p}
\ee
\noindent The operation of $P$ may be identified as a permutation of the
particles `1' and `2' in one dimension. Alternatively, with the
interpretation of $H$ as describing a single-particle system in $N$
dimensions, $P$ is a valid parity transformation in the $N$ dimensional space.
It may be recalled at this point that the transformation,
\bea
&& x_1 \rightarrow x_1 cos \theta + x_2 sin \theta, \nonumber \\
&& x_2 \rightarrow x_1 sin \theta - x_2 cos \theta, \ \
0 \leq \theta \leq 2 \pi,
\eea
\noindent corresponds to parity transformation in two dimensions with
the familiar forms $ x_1 \rightarrow x_1, x_2 \rightarrow - x_2$ or
$ x_1 \rightarrow - x_1, x_2 \rightarrow  x_2$ reproduced by
$\theta=0$ and $\theta=\pi$, respectively. The parity transformation
corresponding to $\theta=\frac{\pi}{2}$, i.e. $x_1 \rightarrow x_2$
and $x_2 \rightarrow x_1$, has been embedded in the $N$
dimensional space for introducing $P$ in Eq. (\ref{2p}) and the choice
of $\theta$ is unique for the Calogero model considered in this article.

The claim is that the non-Dirac-hermitian $H$ is isospectral with the standard
rational Calogero model. The reality of the entire spectra of $H$
could be attributed to an underlying pseudo-hermiticity. A positive-definite
metric $\eta_+$ in the Hilbert space ${\cal{H}}_{\eta_+}$ may be considered:
\be
\eta_+ := e^{- 2 \gamma {\cal{L}}_{12}}, \ \ 
{\cal{L}}_{12} = x_1 p_2 - x_2 p_1, \ \
\gamma \ \in \ R.
\label{metric}
\ee
\noindent The coordinates $x_i$ and the momenta $p_i$ are not hermitian in
${\cal{H}}_{\eta_+}$. A new set of canonical conjugate operators those are
hermitian in the Hilbert space ${\cal{H}_{\eta_+}}$ may be introduced by using
the relation (\ref{obs}) as follows:
\bea
&& X_1 = x_1 \ cosh \phi + i x_2 \ sinh \phi,\nonumber \\
&& X_2 = - i x_1 \ sinh \phi
+ x_2 \ cosh \phi, \ \ X_i=x_i \ for \ i > 2\nonumber \\
&& P_1 = p_1 \ cosh \phi +  i p_2 \ sinh \phi,\nonumber \\
&& P_2 = - i p_1 \ sinh \phi +
p_2 \ cosh \phi, \ P_i= p_i \ for \ i > 2. 
\label{newcor}
\eea
\noindent It may be noted that $L_{12}= X_1 P_2 - X_2 P_1={\cal{L}}_{12}$
is hermitian both in ${\cal{H}}_D$ and ${\cal{H}}_{\eta_+}$. This ensures
that $\eta_+$ defined in Eq. (\ref{metric}) is positive-definite.

The Hamiltonian $H$ can be re-written in terms of ($X_i, P_i$) as,
\be
H =  - \frac{1}{2} \sum_{i=1}^N \frac{\partial^2}{\partial X_i^2}
+ \frac{1}{2} \lambda (\lambda-1) \sum_{i \neq j} X_{ij}^{-2} +
\frac{1}{2} \sum_{i=1}^N  X_i^2,
\ee
\noindent which implies hermiticity of $H$ in ${\cal{H}}_{\eta_+}$.
The non-Dirac-hermitian $H$ can also be mapped to Dirac-hermitian Hamiltonian
$h$ through a similarity transformation,
\bea
h & := & \left ( e^{- \gamma {\cal{L}}_{12}} \right ) \ H \
\left ( e^{ \gamma {\cal{L}}_{12}} \right )\nonumber \\
& = & - \frac{1}{2} \sum_{i=1}^N \frac{\partial^2}{\partial x_i^2}
+ \frac{1}{2} \lambda (\lambda-1) \sum_{i \neq j} x_{ij}^{-2} +
\frac{1}{2} \sum_{i=1}^N  x_i^2.
\label{calo1}
\eea
\noindent Thus, the Hamiltonian $H$ is isospectral with the standard rational
Calogero model $h$. The eigenfunctions $\psi_{\eta_+}$ of $H$ are
related to the eigenfunctions $\psi_D$ of $h$ through the relation,
$ \psi_{\eta_+} = e^{ \gamma {\cal{L}}_{12}} \ \psi_D$. The wave-functions
$\psi_{\eta_+}$ constitute a complete set of orthonormal states in
${\cal{H}}_{\eta_+}$.

A few comments are in order.\\
(i) The rational Calogero model in its original
formulation\cite{csm} has
been first solved for a definite ordering of the particles and then extended
it to the whole of the configuration space in a continuous fashion by using the
underlying permutation symmetry. New states have been found\cite{sae} for $h$ by
considering more general boundary conditions and including singular points/lines
in the configuration space. Thus, with the use of these generalized boundary
conditions, the fact that $H$ is non-singular for $x_1 = x_i, i > 1$ and
$x_2 = x_i,i >2$ has no special significance. However, if singular points/lines
are not included in the configuration space, careful analysis of the
eigenvalue problem of $H$ is required.

(ii) There are  $\frac{N(N-1)}{2}$ numbers of angular-momentum operators
in the $N$-dimensional hyper-spherical coordinate system. More general metric
involving these angular momentum operators can be constructed with new
non-Dirac-hermitian Calogero models admitting entirely real spectra
and unitary time-evolution.

(iii) The construction can be trivially generalized to rational Calogero
models corresponding to other root systems. 

\subsection{Pseudo-hermitian XXZ Spin-chain}

It is a well known fact that non-hermitian quantum spin chains correspond to
two-dimensional classical systems with positive Boltzmann weights.
Examples of non-Dirac-hermitian spin chains are also abundant in the
literature. The list includes XY and XXZ spin chain Hamiltonian with
Dzyaloshinsky-Moriya interaction\cite{electric}, the integrable
chiral Potts model\cite{perk1,albertini}, asymmetric $XXZ$ spin
chains\cite{xxz} and quantum ising spin chain in one dimension\cite{ising}.
Within the context of ${\cal{PT}}$-symmetric theory, non-hermitian spin
chains have been studied in Refs. \cite{piju1,piju,spin}.

A non-Dirac-hermitian XXZ spin-chain in an external complex magnetic field may
be introduced as follows:
\bea
H_A & = & \sum_{i=1}^{N-1} [ \Gamma \left ( e^{w_i - w_{i+1}}
{\cal{S}}_i^+ {\cal{S}}_{i+1}^-
+ e^{-\left ( w_i - w_{i+1} \right )} {\cal{S}}_i^- {\cal{S}}_{i+1}^+ \right )
+ \Delta {\cal{S}}_i^z {\cal{S}}_{i+1}^z \nonumber \\
& + & \left ( A_i cosh w_i - i B_i sinh w_i \right ) {\cal{S}}_i^x
+ \left ( B_i cosh w_i + i A_i sinh w_i \right ) {\cal{S}}_i^y\nonumber \\
& + & C_i {\cal{S}}_i^z ],
\label{diffusion}
\eea
\noindent where ${\cal{S}}_i^{\pm} := {\cal{S}}_i^x \pm i {\cal{S}}_i^y$,
$ \{ \Gamma, \Delta, A_i, B_i, C_i, w_i \} \in R$ and $S_i^{x,y,z}$ are
hermitian in ${\cal{H}}_D$. The non-hermitian
interaction in $H_A$ without the external magnetic field may be interpreted
as arising due to imaginary vector potential. It may be noted that such
imaginary gauge potentials are also relevant in the context of
metal-insulator transitions or depinnning of flux lines from extended
defects in type-II superconductors\cite{hn}. A Hamiltonian resembling
the random-hopping model of Ref. \cite{hn} can be obtained from $H_A$
by using a hard-core boson representation and mapping it to a
non-hermitian quadratic form of bosonic operators with nearest-neighbour
interactions\cite{piju1}. The inclusion of the complex magnetic field
is justified, since it may shed light on the nature of ordinary second order
phase transitions as in the case of popular Yang-Lee model\cite{yl}.
As discussed below, $H_A$ reduces to the asymmetric $XXZ$ model\cite{xxz}
that arises in the context of two species reaction-diffusion
processes and Kardar-Parisi-Zhang-type growth phenomenon. 

The parity and the time-reversal transformations involving spin-operators
are defined as,
\be
P: \vec{S}_i \rightarrow \vec{S}_i, \ \
T: \vec{S}_i \rightarrow - \vec{S}_i, \ \ i \rightarrow -i.
\ee
\noindent It may be noted that spin being an axial vector does not change
sign under the parity operation. The Hamiltonian $H_A$ is not invariant under
the combined operation
of $PT$. However, the Hamiltonian $H_A$ with $A_i=0=B_i \ \forall \ i$ is
invariant under an anti-linear ${\cal{PT}}$ transformation, where 
${\cal{T}}: i \rightarrow -i$ and ${\cal{P}}$ is defined as a discrete
symmetry in the spin-space with its actions on the spin operators as
follows:
\bea
S_i^x & \rightarrow & \bar{S}_i^x = S_i^x cos \theta +
S_i^y sin \theta\nonumber\\
S_i^y & \rightarrow & \bar{S}_i^y = S_i^x sin \theta -
S_i^y cos \theta\nonumber\\
S_i^z & \rightarrow & \bar{S}_i^z = S_i^z, \ \ 0 \leq \theta \leq 2 \pi.
\label{pipi}
\eea
\noindent The discrete transformation ${\cal{P}}$ is similar to a non-standard
parity transformation in three dimensions involving the position co-ordinates.
The ${\cal{PT}}$ symmetry can be promoted to be the symmetry of $H_A$ with
non-vanishing $A_i$ and  $B_i$ for a fixed $\theta$, provided these parameters
are related to each other through the relations:
\be
\frac{B_i}{A_i} = tan \frac{\theta}{2}, \ \ \forall \ i.
\ee
\noindent The Hamiltonian $H_A$ may be invariant under a more general
anti-linear transformation for arbitrary $A_i$ and $B_i$ which
is not known at this point. 

The Hamiltonian $H_A$ is hermitian in ${\cal{H}}_{\eta_+}$ with the metric:
\be
\eta_+ := \prod_{i=1}^N e^{-2 \gamma_i S_i^z}.
\ee
\noindent A set of spin-operators $T_{x,y,z}$ may be introduced which are
hermitian in ${\cal{H}}_{\eta_+}$:
\bea
T_i^x & := & cosh w_i \ {\cal{S}}_i^x +
i sinh w_i \ {\cal{S}}_i^y\nonumber \\
T_i^y & := & - i sinh w_i \ {\cal{S}}_i^x +
cosh w_i \ {\cal{S}}_i^y\nonumber \\
T_i^z & := & {\cal{S}}_i^z.
\eea
\noindent Consequently, $H_A$ can be re-written as,
\be
H_A = \sum_{i=1}^{N-1} \left [ \Gamma \left ( T_i^+ T_{i+1}^-
+ T_i^- T_{i+1}^+ \right ) + \Delta T_i^z T_{i+1}^z
+ A_i T_i^x + B_i T_i^y + C_i T_i^z \right ],
\ee
\noindent showing its hermiticity in ${\cal{H}}_{\eta_+}$,
where $T_i^{\pm} := T_i^x \pm i T_i^y$.
The Hamiltonian $H_A$ can be mapped to a Dirac-hermitian Hamiltonian, 
\bea
h & := & ( \eta_+^{\frac{1}{2}} H_A \eta_+^{-\frac{1}{2}} )\nonumber \\
& = & \sum_{i=1}^{N-1}  \left [ \Gamma \left ( {\cal{S}}_i^x {\cal{S}}_{i+1}^x +
{\cal{S}}_i^y {\cal{S}}_{i+1}^y \right ) +
\Delta {\cal{S}}_i^z {\cal{S}}_{i+1}^z +
A_i {\cal{S}}_i^x + B_i {\cal{S}}_i^y + C_i {\cal{S}}_i^z \right ],
\label{xxhermi}
\eea
\noindent implying that both $H_A$ and $h$ have entirely real spectra.

A few comments are in order at this point.\\
(i) A typical choice for $w_k$ as $w_k =w - (k-1) \phi$ leads to a
site-independent global phase factor $e^{\pm \phi}$ in lieu of
$e^{\pm (w_i - w_{i+1})}$ and $H_A$ reduces to asymmetric XXZ
Hamiltonian that has been studied in the
literature\cite{xxz} in the context of two species reaction-diffusion
processes and Kardar-Parisi-Zhang-type growth phenomenon. Although the
transformation that maps non-hermitian asymmetric $XXZ$ Hamiltonian
to a hermitian Hamiltonian is known in the literature\cite{xxz}, the
realization of pseudo-hermiticity is new. Thus, with the discovery of the
pseudo-hermiticity of $H_A$, it may be used to describe unitary time
evolution in ${\cal{H}}_{\eta_+}$.

(ii) With the choice of $w_i \equiv w \ \forall \ i$ in Eq. (\ref{diffusion}),
a symmetric $XXZ$ spin-chain Hamiltonian in an external complex
magnetic field may be constructed, 
\bea
H_S & = & \sum_{i=1}^{N-1}  [ \Gamma \left ( {\cal{S}}_i^x {\cal{S}}_{i+1}^x +
{\cal{S}}_i^y {\cal{S}}_{i+1}^y \right ) +
\Delta {\cal{S}}_i^z {\cal{S}}_{i+1}^z +
\left ( A_i cosh w - i B_i sinh w \right ) {\cal{S}}_i^x\nonumber \\
& + & \left ( B_i cosh w + i A_i sinh w \right ) {\cal{S}}_i^y
+ C_i {\cal{S}}_i^z ],
\label{symxxz}
\eea
\noindent which is non-Dirac-hermitian, but, hermitian in
${\cal{H}}_{\eta_+}$. The equivalent Dirac-hermitian Hamiltonian
$h := ( \eta_+^{\frac{1}{2}} H_S \eta_+^{-\frac{1}{2}} )$
is still given by Eq. (\ref{xxhermi}).

The Hamiltonian $h$ has several integrable limits. Consequently,
$H_A$ and $H_S$ are also integrable in these limits with entirely real
spectra and unitary time-evolution. For example, $h$ reduces to a
transverse-field Ising model for $\Gamma=B_i=C_i=0, A_i=A \ \forall \ i$
and both $h$ and $H_S$ have been studied in some detail\cite{piju} for
this limiting case. For $\Delta=0, A_i=0, B_i=0 \ \forall \ i$, $h$ reduces to
an XX model in a transverse magnetic field and  is exactly
solvable\cite{lsm,xx}. Although $H_S$ is hermitian in ${\cal{H}}_D$ for this
choice of the parameters, $H_A$ is non-hermitian. Thus, the non-hermitian $H_A$
is exactly solvable and has an equivalent description in terms of
a hermitian XX model in an external magnetic field. For the following
choice of the parameters,
\bea
&& \Gamma=1, \Delta= cosh q, C_1 = -C_N = - sinh q,\nonumber \\
&& A_i=B_i=0 \ \forall \ i; C_i = 0, i=2, 3, \dots, N-1,
\eea
\noindent $h-\Delta$ reduces to an $SU_q(2)$ invariant\cite{qg}
integrable\cite{inami} spin-chain Hamiltonian. The XXZ spin-chain
with $Sl_2$ loop symmetry\cite{deguchi} may also be obtained as a
limiting case. The corresponding non-hermitian Hamiltonian $H_A$ is
also integrable and allows an unitary description.

(iii) A more general ${\cal{PT}}$-symmetric Hamiltonian may be introduced
which contains $H_A$ as a special case. In particular,
\bea
\tilde{H}_A & = & \sum_{i=1}^{N-1} \left [ \Gamma \left ( \gamma_{i,i+1} 
{\cal{S}}_i^+ {\cal{S}}_{i+1}^- + \delta_{i,i+1} 
{\cal{S}}_i^- {\cal{S}}_{i+1}^+ \right )
+ \Delta {\cal{S}}_i^z {\cal{S}}_{i+1}^z \right ] \nonumber \\
& + & \sum_{i=1}^{N-1} \left [ \left ( \alpha_i^R +
i \alpha_i^I \right ) {\cal{S}}_i^x
+ \left ( \beta_i^R + i \beta_i^I \right ) {\cal{S}}_i^y
+ C_i {\cal{S}}_i^z \right ],
\label{1diffusion}
\eea
\noindent is invariant under the ${\cal{PT}}$ transformation provided that,
for a fixed $\theta$, the real parameters $\alpha_i^R, \alpha_i^I,
\beta_i^R, \beta_i^I$ satisfy the equations,
\be
\frac{\beta_i^R}{\alpha_i^R} = - \frac{\alpha_i^R}{\beta_i^I}=
tan (\frac{\theta}{2}) \ \ \forall \ i.
\ee
\noindent It may be noted that there are no restrictions on the real
parameters $\gamma_{i,i+1}$ and $\delta_{i,i+1}$ in order to ensure the
${\cal{PT}}$ symmetry. Further, $\alpha_i^{R,I}$ and $\beta_i^{R,I}$ are 
independent of the parameters $\gamma_{i,i+1}$ and $\delta_{i,i+1}$
in $\tilde{H}_A$. The Hamiltonian $H_A$ is reproduced with the choice of
the parameters:
\bea
&& \gamma_{i,i+1}= e^{w_i - w_{i+1}}, \ \ \delta_{i,i+1}=
 e^{-(w_i - w_{i+1})}\nonumber \\
&&\alpha_i^R=A_i cosh w_i, \ \alpha_i^I= - B_i sinh w_i,\nonumber \\
&& \beta_i^R=B_i cosh w_i, \ \beta_i^I=  A_i sinh w_i.
\eea
\noindent The region in the parameter-space of $\tilde{H}_A$ for which
${\cal{PT}}$ is unbroken must contain the region defined by the above
equations. A concrete investigation to find the region of unbroken
${\cal{PT}}$ symmetry for $\tilde{H}_A$ is desirable.

(iv) The general prescription given in this article may be used to construct
models of non-Dirac-hermitian spin-chain with long-range interaction.
For example, the non-Dirac-hermitian spin-chain Hamiltonian,
\be
H = \pm \sum_{i < j} \frac{\vec{T}_i \cdot {\vec{T}_j}}{ 2 sin^2
\frac{\pi}{N} (i-j) },
\ee
\noindent is isospectral with the celebrated Haldane-Shastry\cite{hs}
model. The equivalent hermitian Hamiltonian in ${\cal{H}}_D$
may be obtained as,
\be
h := \rho H \rho ^{-1}
= \pm \sum_{i < j} \frac{\vec{\cal{S}}_i \cdot {\vec{\cal{S}}_j}}{ 2 sin^2
\frac{\pi}{N} (i-j) }.
\ee
\noindent It may be noted that,
in general, eigenstates of $h$ and $H$ are different. However, with
proper identification of physical observables in ${\cal{H}}_{\eta_+}$
through Eq. (\ref{obs}), different correlation functions of the quantum
systems governed by $H$ and $h$ are identical.

\section{Conclusions}

A class of non-Dirac-hermitian many-particle quantum systems admitting
entirely real spectra has been presented. The time-evolution of these
systems is guaranteed to be unitary with the modified inner-product in
the Hilbert space involving the pre-determined metric. These quantum systems
are isospectral with known Dirac-hermitian quantum systems and are exactly
solvable. In fact, several previously studied non-Dirac-hermitian quantum
systems involving transverse ising-chain\cite{piju}, asymmetric XXZ
spin-chain\cite{piju1,xxz} belong to this class of pseudo-hermitian quantum
system. New exactly solvable interesting pseudo-hermitian many-particle quantum
systems involving rational Calogero model, XXZ spin-chain and Haldane-Shastry
spin-chain have also been constructed. Many other physically relevant
pseudo-hermitian quantum system involving  Swanson model\cite{swan},
Dicke model, Lipkin model, quadratic boson/fermion form  etc. are
described in Ref. \cite{piju1}.

The general approach that has been followed in the construction of these
quantum systems is the following. The basic canonical commutation relations
defining these systems have been realized in terms of non-Dirac-hermitian
operators which are hermitian with respect to the modified inner product in the
Hilbert space involving the pre-determined metric. Consequently, appropriate
combinations of these operators result in a very large number of
pseudo-hermitian quantum systems. The construction in this article is purely
mathematical. Physical realizations of such models are desirable.


\end{document}